\documentclass[pre%
,twocolumn%
,superscriptaddress%
,floats% ,dcolumn%
,aps%
,amssymb%
]{revtex4}
\usepackage{amsmath}%
\begin{document}
\title{Obtaining Stiffness Exponents from Bond-diluted Lattice Spin Glasses} 
\date{\today}
\author{S. Boettcher and S. E. Cooke}  
\email{www.physics.emory.edu/faculty/boettcher}
\affiliation{Physics Department, Emory University, Atlanta, Georgia
30322, USA}  

\begin{abstract} 
Recently, a method has been proposed to obtain accurate predictions
for low-temperature properties of lattice spin glasses that is
practical even above the upper critical dimension, $d_c=6$. This
method is based on the observation that bond-dilution enables the
numerical treatment of larger lattices, and that the subsequent
combination of such data at various bond densities into a finite-size
scaling Ansatz produces more robust scaling behavior. In the present
study we test the potential of such a procedure, in particular, to
obtain the stiffness exponent for the hierarchical Migdal-Kadanoff
lattice. Critical exponents for this model are known with great
accuracy and any simulations can be executed to very large lattice
sizes at almost any bond density, effecting a insightful comparison
that highlights the advantages -- as well as the weaknesses -- of this
method. These insights are applied to the Edwards-Anderson model in
$d=3$ with Gaussian bonds.

\hfil\break  PACS number(s): 
05.50.+q%Lattice theory and statistics (Ising, Potts, etc.)
%, 05.40.-a%Fluctuation phenomena, random processes, noise, and Brownian motion
%, 64.70.Pf%Glass transitions, 
%, 64.60.Cn%Order-disorder transformations; statistical mechanics of model systems
, 75.10.Nr%Spin-glass and other random models
, 02.60.Pn%Numerical optimization
.
\end{abstract} 
\maketitle

\section{Introduction}
\label{intro}
The exploration of low-temperature properties of disordered systems
remains an important and challenging problem~\cite{YoungBook}. Systems
in this class possess a low-temperature glassy state with a glass
transition at some temperature $T_g>0$. They are characterized by a
complex (free-)energy landscape in configuration
space~\cite{cnls,SD,BS} with a hierarchy of valleys and barriers whose
multi-modal structure impedes the progression of any dynamics towards
equilibration, causing tantalizing phenomena, such as trapping and
jamming on intermediate time scales, and aging on long time scales. An
understanding of such systems is of paramount importance as these
phenomena are observed for a large class of materials as well as for
biological systems~\cite{Dagstuhl}.

The paradigmatic model for the study of such phenomena is the Ising
spin glass, either on a finite-dimensional lattice (Edwards-Anderson
model, EA~\cite{EA}) or some other random network structure. Disorder
effects arise typically via quenched random bonds, random local
fields, or merely the randomness of the network itself, each can lead
to conflicting constraints which leave variables frustrated. It is
believed that a proper understanding of static and dynamic features of
EA may aid a description of the unifying principles expressed in the
wider class of realistic problems~\cite{F+H}.

Unfortunately, after 30 years of research there is still not consensus
on whether the subtle mean-field picture for Ising spin glasses
derived long ago~\cite{SK,ParisiRSB,MPV} has any relevance for
real-world materials at low temperatures~\cite{FH}. Progress towards
such an understanding is slow due to the peculiar structure of the
problem at hand. The intricate multi-modal low-energy landscape puts
the computational effort needed to determine thermodynamic observables
usually into the class of NP-hard combinatorial optimization
problems~\cite{Barahona} know from computer science, for which worst
case computational costs increase faster than any power of the system
size. (Interestingly, it is in this area of computer science itself
where the mean-field theory at $T=0$ has had a most significant impact
so far ~\cite{MPZ}.)

Most insights into finite-dimensional systems has thus been gained
through alternative computational approaches to elucidate
low-temperature properties. Aside from methods designed to expedite
some thermodynamically correct algorithm, such as parallel
tempering~\cite{SimTemp} or the waiting time method~\cite{Dall01}, one
focus area has been the use of optimization heuristics directed toward
fully enumerating ground-state configurations by any
means~\cite{Dagstuhl,Pal,Hartmann,Palassini,eo_prl,Middleton}. These
ground states should provide the basis for the thermodynamic behavior
of the system at and near $T=0$. But due to the NP-hardness, even
heuristic methods become unreliable when system sizes exceed about
$\approx10^3$ variables, often too small to draw safe conclusions or to
sufficiently discriminate between theoretical ideas~\cite{KM,PY}.

This fact is illustrated by the determination of the stiffness
exponent, often labeled $y$ or $\theta$~\cite{BM,F+H}, a fundamental
quantity assessing low-temperature energy fluctuations: a positive
value of $y$, as found in EA for $d\geq3$, denotes the increase in the
energetic cost (i. e. ``stiffness'') accompanying a growing number of
variables perturbed from their position in the ground state. The
rise in energetic penalty paid for stronger disturbances signals the
presence of an ordered state. In turn, for systems with $y\leq0$ such
order is destabilized by arbitrarily small fluctuations.

In this paper, we will extract the stiffness exponent from the
response induced through defect-interfaces~\cite{BM}. These can be
created by fixing the spins along the two faces of an open boundary in
one lattice direction. The ground state configuration with energy
$E_0$ of an instance is first determined for a random fixing of those
boundary spins, then the energy $E_0'$ is obtained for the same
instance and the same fixing, but with all spins reversed on one of
the faces. Hence, the interface energy $\Delta E=E_0'-E_0$ created by
the perturbation on the boundary is sampled, and its distribution
$P(\Delta E)$ determined. If a system is glassy, the typical energy
scale involved, here represented by the width of the distribution,
$\sigma(\Delta E)=\sqrt{\langle\Delta E^2\rangle-\langle\Delta
E\rangle^2}$, should grow with the size of the perturbation, say, the
linear extend of the boundary, $L$, as~\cite{BM}
\begin{eqnarray}
\sigma(\Delta E)\sim L^y.
\label{yeq}
\end{eqnarray}
Accurate determination of this exponent in $d=3$ has long been
elusive, with values given between $y\approx0.19$~\cite{BM,Hartmann}
to $\approx0.27$~\cite{CBM}. While it was save to say that $y>0$, its
value could be at best given as $y_3=0.2(1)$. In terms of sorting out
theoretical models, this is too inaccurate to draw solid
conclusions~\cite{PY}.

\begin{figure}
\vskip 2.1in \includegraphics{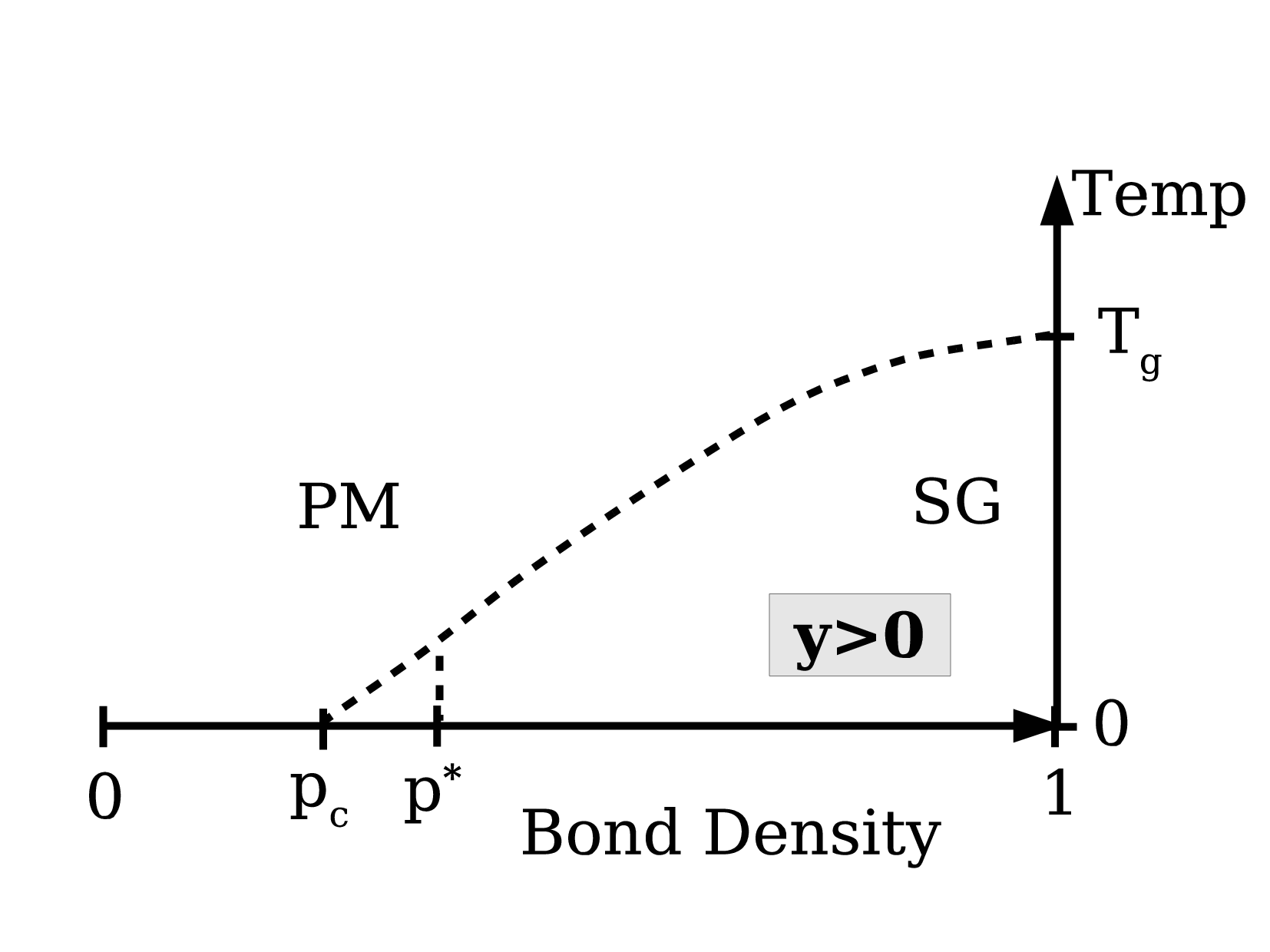}
\caption{Phase diagram for bond-diluted spin glasses at low
  temperatures. Lattice spin glasses of dimension $d\geq3$ possess a
  spin glass phase (SG) below the glass transition temperature
  $T_g>0$. This phase should persist even for diluted lattices, as
  long as the percolating cluster of connected spins is compact,
  i.e. for bond densities $p$ sufficiently above
  $p_c$~\protect\cite{BF,stiff1,stiff2}. Below $p_c$, the collection
  of small clusters can only respond paramagnetically (PM) on long
  lengthscales for any $T\geq0$. Depending on the details of the bond
  distribution, the $T=0$ transition occurs at $p^*=p_c$ (continuous
  bonds) or at $p^*>p_c$ (discrete bonds), respectively determining the slope
  of the phase boundary $T_g(p)$ at $p^*$.}
\label{phaseplot}
\end{figure}

This limited accuracy originates with trying to fit data over too
small a range of system sizes $L$. Scaling in this range is further
beset by slowly decaying corrections in $L$ which make it hard to
decide even where scaling sets in~\cite{DM,HY}. The remedy proposed
then is to possibly extend scaling by considering {\it bond-diluted}
lattices~\cite{stiff1,stiff2,Chak}. As has been argued already in
Ref.~\cite{BF}, as soon as the percolation window for bond densities
$p>p_c$ has been exceeded at some lattice size, the long-range
properties of the giant component -- and hence, of the spin glass
defined on it -- are essentially compact, representing the $T=0$
fixed-point of the fully connected lattice. This implies that there
the exponent $y$ is independent of the bond density $p$. These
features of bond-diluted glasses are summarized in the phase diagram
in Fig.~\ref{phaseplot}. The independence of $y$ from $p$ appears to
hold also for Gaussian bonds in $d=2$~\cite{BH}, where $y<0$ (no spin
glass phase) and universality with respect to bond disorder is
violated~\cite{HY,AMMP,BKM}.

While we focus on the defect energy at $T=0$ here, bond-dilution may
also be an effective means to study other observables~\cite{Jorg}. For
determining $T=0$ properties, this approach makes the treatment of
larger lattices sizes practical in as much as exact algorithms can be
devised that allow the elimination of a large number of variables,
whose state becomes entrained  to other variables in a predictable way
in the ground state~\cite{MKpaper}. In this way, systems with more
than $10^6$ variables have been ``reduced'' to remainder graphs
consisting of no more than a few hundred variables that are amenable
to standard optimization methods~\cite{stiff1}. Clearly, though, due
to those almost trivial variables, the information contend about the
asymptotic behavior captured by such graphs may be limited, and a
price has to be paid through extended transient behavior. As we hope
to demonstrate here, in the end much is gained in this method,
although at the extremely large lattice sizes that can be reached with
the Migdal-Kadanoff approximation used here, diminished returns are
obtained because of a lack of knowledge on how scaling corrections
vary with bond density.

As we will recount in Sec.~\ref{defect}, Refs.~\cite{BBF,BF} suggest
to generalize Eq.~(\ref{yeq}) to
\begin{eqnarray}
\sigma(\Delta E)_{L,p}\sim \xi(p)^{y_P}\left(\frac{L}{\xi(p)}\right)^y
f\left(\frac{L}{\xi(p)}\right),
\label{extendedyeq}
\end{eqnarray}
for $L\gg1$ {\it and} $\xi(p)\sim(p-p^*)^{-\nu^*}\gg1$. The scaling
function $f$ was chosen to be constant for $L\gg\xi(p)$. The limit
$p\to p^*$ towards the $T=0$-transition between the spin glass and
paramagnetic regime (see Fig.~\ref{phaseplot}) is interesting in its
own right, and will be investigated further in
Ref.~\cite{Marchetti}. Yet, to elucidate properties of the glassy
regime, the ``window of opportunity'' for our method appears to be at
intermediate bond densities: $p$ has to be sufficiently smaller than
unity for our reduction algorithm to be efficient, but also
sufficiently {\it above} $p^*$ to attain system sizes
$L\gg\xi(p)$. Analyzing our numerical data here suggests that in this
window the condition $\xi(p)\sim(p-p^*)^{-\nu^*}\gg1$ assumed in
Eq.~(\ref{extendedyeq}) does not hold, making $\xi(p)$ a more general
function of $p$ with unknown corrections to its singular part.

Operationally, we thus propose a naive Ansatz for a collapse of the
numerical data valid for the asymptotic regime $L\gg\xi(p)\sim1$ only,
\begin{eqnarray}
\sigma(\Delta E)= f(\infty)~x^y\qquad\left[x=L(p-p^*)^{\nu^*}\right].
\label{fiteq}
\end{eqnarray}
Since the numerically accessible data appears to violate $\xi(p)\gg1$,
the parameters $p^*$ and $\nu^*$ here merely facilitate the data
collapse by fitting a more general function $\xi(p)$, and can not be
expected to yield accurate predictions for the critical values in
Eq.~(\ref{extendedyeq}). The successful implementation of this Ansatz
for the Migdal-Kadanoff hierarchical lattice here lends credibility to
the findings for $y_d$ of the EA in Refs.~\cite{stiff1,stiff2}.

Finally, it is remarkable that this approach works substantially
better for a discrete $\pm J$ bond distribution than for Gaussian
bonds, for which scaling corrections due to a small $\xi(p)$, and even
due to finite-$L$ transients, are far more significant. Just as for
the $\pm J$ data, the similarity between our Migdal-Kadanoff and our
EA data with Gaussian bonds presented here is striking. While this
issue eventually deserves more thorough investigation, we can speculate
on the origin of these strong transients. One may remember that in
$d=1$, $\sigma(\Delta E)\sim L^{-1}$ for continuously distributed
bonds with finite $P(0)$, while $\sigma(\Delta E)\sim L^0$ for $\pm J$
bonds. The interface may settle on the extremely weakest within a more
widely distributed set of continuous bonds, while for $\pm J$ it {\it
must} break a bond of order unity. Similarly, on short ranges in
higher dimensional lattices, particularly dilute ones, with a Gaussian
distribution peaked at $J=0$, the violation of heavy bonds can be
deferred {\it initially} that an extended defect eventually demands on
larger scales $L$. This naive argument is supported by the fact that
at equal $p$ and $L$, defect energies $\sigma(\Delta E)$ are typically
somewhat smaller for Gaussian than for $\pm J$ bonds, as will be seen
in Fig.~\ref{defectscal}.

In the next section we introduce the Migdal-Kadanoff hierarchical
lattice, followed in Sec.~\ref{defect} by a discussion of the scaling
arguments leading to Eq.~(\ref{extendedyeq}) and to our Ansatz in
Eq.~(\ref{fiteq}) used to obtain the stiffness exponent on
bond-diluted lattices. Sec.~\ref{numerics} contains our discussion of
the Migdal-Kadanoff data, followed by the application of the gained
insights to the diluted EA with Gaussian bonds in
Sec.~\ref{Gauss}. Finally, Sec.~\ref{conclusion} contains our
conclusions.

\section{Migdal-Kadanoff Hierarchical Lattice}
\label{MKlat} 
\begin{figure}
\vskip 3.7in \includegraphics{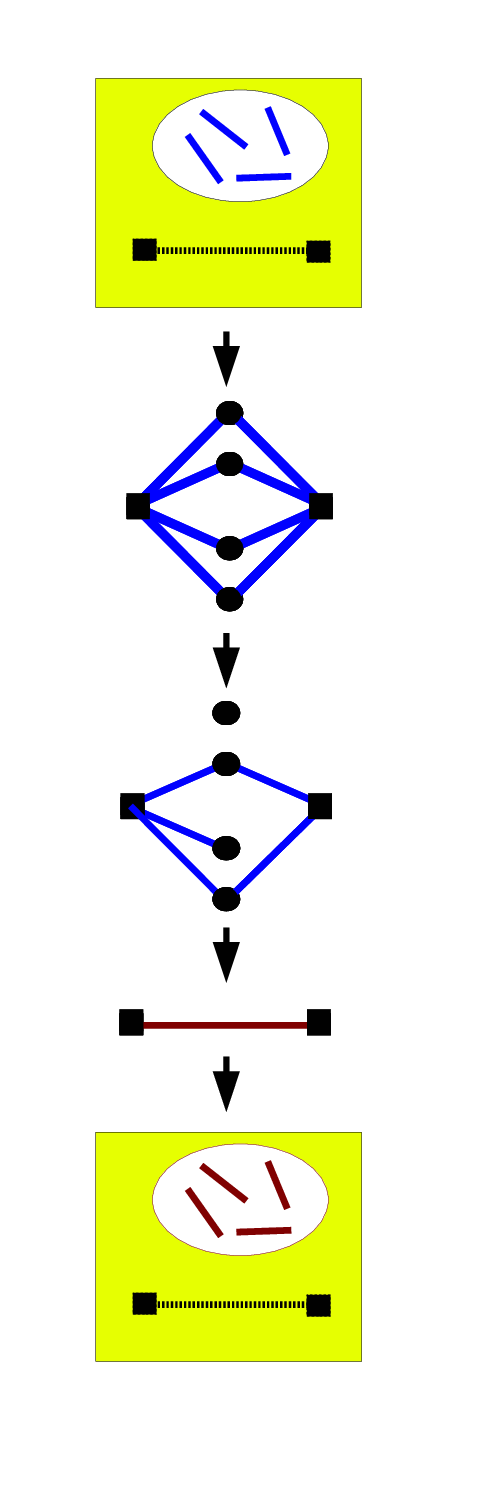}
\caption{Diagram for the recursive algorithm to calculate spin glasses
  on hierarchical lattices. Drawing bonds $J_{I-1}$ randomly from a
  sufficiently large pool $P_{I-1}(J)$ at generation $I-1$ (top), a
  lattice of generation $I$ is formed, then bond-diluted to density
  $p$, and finally ``reduced'' \protect\cite{MKpaper} to an effective
  bond $J_I$ between the root spins (squares), which is add to the
  pool $P_I(J)$ (bottom). }
\label{hierlat}
\end{figure}

To test the scaling Ansatz in Eq.~(\ref{fiteq}) for the stiffness
exponent, we consider here the bond-diluted hierarchical lattice (see
Fig.~\ref{hierlat}), obtained in the Migdal-Kadanoff bond-moving
scheme \cite{MK} for low-dimensional spin glasses. These lattices have
a simple recursive yet geometric structure and are well-studied
\cite{Kirkpatrick,SY,BF,DM}. Most importantly, their ground states can
be obtained in polynomial time at any bond density, and we can discuss
our results independent of any systematic bias introduce by a
subsequent optimization process that may be required for more
complicated models like EA~\cite{MKpaper}. A most interesting property
of these lattices is the curious fact that the scaling of its defect
energy distribution, giving rise to the stiffness exponent, behaves in
all respects very similar to that measured for actual
three-dimensional lattices~\cite{SY,BM,F+H}.

To generate a hierarchical lattice, starting from generation $I=0$
with a single link, at each subsequent generation $I$ all links from
$I-1$ are replace with a new subgraph, as described in
Fig.~\ref{hierlat}. The structure of the subgraph arises from the
Migdal-Kadanoff bond-moving scheme in $d$ dimensions, and has $2^d=8$
links for $d=3$ here. Thus, a hierarchical lattice of generation $I$
has $(2^d)^I$ links (if undiluted), thus corresponding to a
$d$-dimensional lattice of ``length'' $L=2^I$ and
$n=2+2^{d-1}(L^d-1)/(2^d-1)=O(L^d)$ vertices. While the average
connectivity is $\sim4-2^{2-d}$, the two root-vertices from generation
$I=0$ themselves obtain in generation $I$ a connectivity of
$\sim2^{(d-1)(I-1)}$. In turn, $\sim2^{dI-1}$ vertices, 7 in 8 for
$d=3$, are only two-connected.

The diluted hierarchical lattice percolates when there is a path
between the two root-vertices, representing the boundaries of the
system. This notion leads to a simple recursion relation for the
percolation threshold by counting the weights of all diluted subgraphs
from Fig.~\ref{hierlat} that percolate, i.~e. connect right and left
vertex. In $d=3$ one gets
\begin{eqnarray}
p_{I+1}=4p_I^2-6p_I^4+4p_I^6-p_I^8,
\label{perceq}
\end{eqnarray}
which has a non-trivial stationary point at $p_c=0.2818376366$. There
can not be long-range correlated behavior, such as spin glass
ordering, for any $p\leq p_c$.  It has been pointed out by
Ref.~\cite{BF} that a spin glass on a hierarchical lattice with $\pm
J$-bonds exhibits a critical transition between a paramagnetic and a
spin glass phase for bond densities at $p^*=0.31032$. While below
$p_c$ disconnected clusters clearly prevail and prevent long-range
correlations, even for $p_c<p<p^*$ such correlations remain suppressed
due to the cooperative behavior in the bond structure pervasive in the
lattice, leading to cancellations that additionally disconnects
subgraphs at some higher level of the hierarchy. In contrast, for a
continuous bond distribution, such as the Gaussian bonds discussed
below, any such cancellations would be unlikely, leading in this case
to $p^*=p_c$.

\section{Defect-Energy Scaling for $p\to p^*$} 
\label{defect}
Following the discussion in Refs.~\cite{BBF,BF}, for bond-diluted
lattices at $p\to p^*$ we have to generalize the scaling relation in
Eq.~(\ref{yeq}) for the defect energy to
\begin{eqnarray}
\sigma(\Delta E)_{L,p}\sim {\cal Y}(p) L^y
f\left(\frac{L}{\xi(p)}\right),
\label{sigmaeq}
\end{eqnarray}
where ${\cal Y}\sim(p-p^*)^t$ is an effective surface tension and
$\xi(p)\sim(p-p^*)^{-\nu^*}$ is the correlation length for the
cross-over into glassy behavior. Note that Eq.~(\ref{sigmaeq})
requires both, $L$ and $\xi$, to be large compared with the unit
lattice spacing to avoid further scaling corrections. The scaling
function $f(x)$ is defined to be constant for large argument,
$L\gg\xi\gg1$.

For $\xi\gg L\gg1$, Eq.~(\ref{sigmaeq}) requires that $f(x)\sim
x^{\mu}$ for $x\to0$ to satisfy either the vanishing of $\sigma$ with
$L$ in case of $p^*=p_c$ (Gaussian bonds), or its scale invariance at
a $p^*>p_c$ ($\pm J$ bonds). Clearly, due to the tenuous fractal
nature of the percolating cluster at $p^*=p_c$, no long-range order
can be sustained, defects possess a vanishing interface, and one may
expect that
\begin{eqnarray}%
\sigma(\Delta E)_{L,p^*}\sim L^{y_P},%
\label{yPeq}%
\end{eqnarray}%
where $y_P\leq0$~\cite{BBF}. To cancel the $p$-dependence at $p=p^*$,
Eq.~(\ref{sigmaeq}) requires $y+\mu=y_P$ and $t+\mu\nu=0$,
i.~e. $t=\nu y+\phi$, setting $\phi=-\nu y_P$~\cite{BBF}. As a result,
we obtain Eq.~(\ref{extendedyeq}).  In contrast, for $p^*>p_c$, the
percolating cluster appears compact on scales $L\gg(p^*-p_c)^{-\nu}$,
and $\sigma$ can not vanish for further increasing $L$. Yet, neither
can $\sigma$ increase at $p^*$, by definition. Hence, $\sigma$ remains
scale invariant, and correspondingly $\phi$ and $y_P$ vanish in
Eqs.~(\ref{extendedyeq}) and~(\ref{yPeq}) for $\pm J$ bonds.

In the case of continuous bond-distributions $P(J)$ with finite
$P(0)$, where $p^*=p_c$, spin glass ordering can only be noticed on
scales at least as large as the correlation length
$\xi\sim(p-p_c)^{-\nu}$ associated with percolation, suggesting
$\nu^*=\nu$. In case of $\pm J$-bonds, the cooperative cancellations
between discrete bonds mentioned above further weaken order, leading
to $\nu^*>\nu$ for the $T=0$ glass transition at $p^*$. Accordingly,
Ref.~\cite{BF} finds $\nu=1.2274$ and $\nu^*=1.5373$ for the
hierarchical lattice in $d=3$.

Finally, at the cross-over $\xi\sim L$, where the range $L$ of the
energy excitations $\sigma(\Delta E)$ reaches the percolation length
and spin glass order ensues, Eq.~(\ref{sigmaeq}) yields
\begin{eqnarray}
\sigma(\Delta E)_{\xi(p),p}\sim\left(p-p^*\right)^t\xi^yf(1)
\sim\left(p-p^*\right)^{\phi}.
\label{phieq}
\end{eqnarray}
One can associate a characteristic temperature with this cross-over by
$\beta\sigma(\Delta E)_{\xi(p),p}\sim 1$; for temperatures above this
$T=1/\beta$, thermal fluctuation destroy spin glass order. This
suggests a relation between bond density and the glass transition
temperature:
\begin{eqnarray}
T_g(p)\sim\left(p-p^*\right)^{\phi}.
\label{Tgeq}
\end{eqnarray}
Eq.~(\ref{Tgeq}) defines $\phi$ as the ``thermal-percolative
cross-over exponent''~\cite{BBF}, which specifies the details of the
phase boundary near $p_c$ (or $p^*$) in Fig.~\ref{phaseplot}. For the
continuous distribution, where $p^*=p_c$, $\phi$ would be in general
nontrivial, while for $\pm J$ at $p^*>p_c$ it appears that $\phi=0$,
indicating a jump in the phase boundary $T_g(p)$ at $p^*$.

The exponent $\phi$ could be of importance, since it may be
experimentally accessible~\cite{Beckman,Maletta} while the relation
$\phi=-y_P\nu$ provides a simple computational determination in terms
of Eq.~(\ref{yPeq}) and the well-known percolation exponents $\nu$.
This connection will be considered in a forthcoming
publication~\cite{Marchetti}.

In the following, we will explore some of these relations numerically
for the hierarchical lattice, for which large sizes $L$ can be
obtained. But the central purpose of this paper is to probe
Eq.~(\ref{fiteq}), which has been used in Refs.~\cite{stiff1,stiff2}
to provide accurate predictions for the stiffness exponents $y_d$,
fundamental for describing low-temperature excitations in spin
glasses.

\section{Numerical Results for Hierarchical Lattices}
\label{numerics}
\begin{figure}
\vskip 4.4in \includegraphics{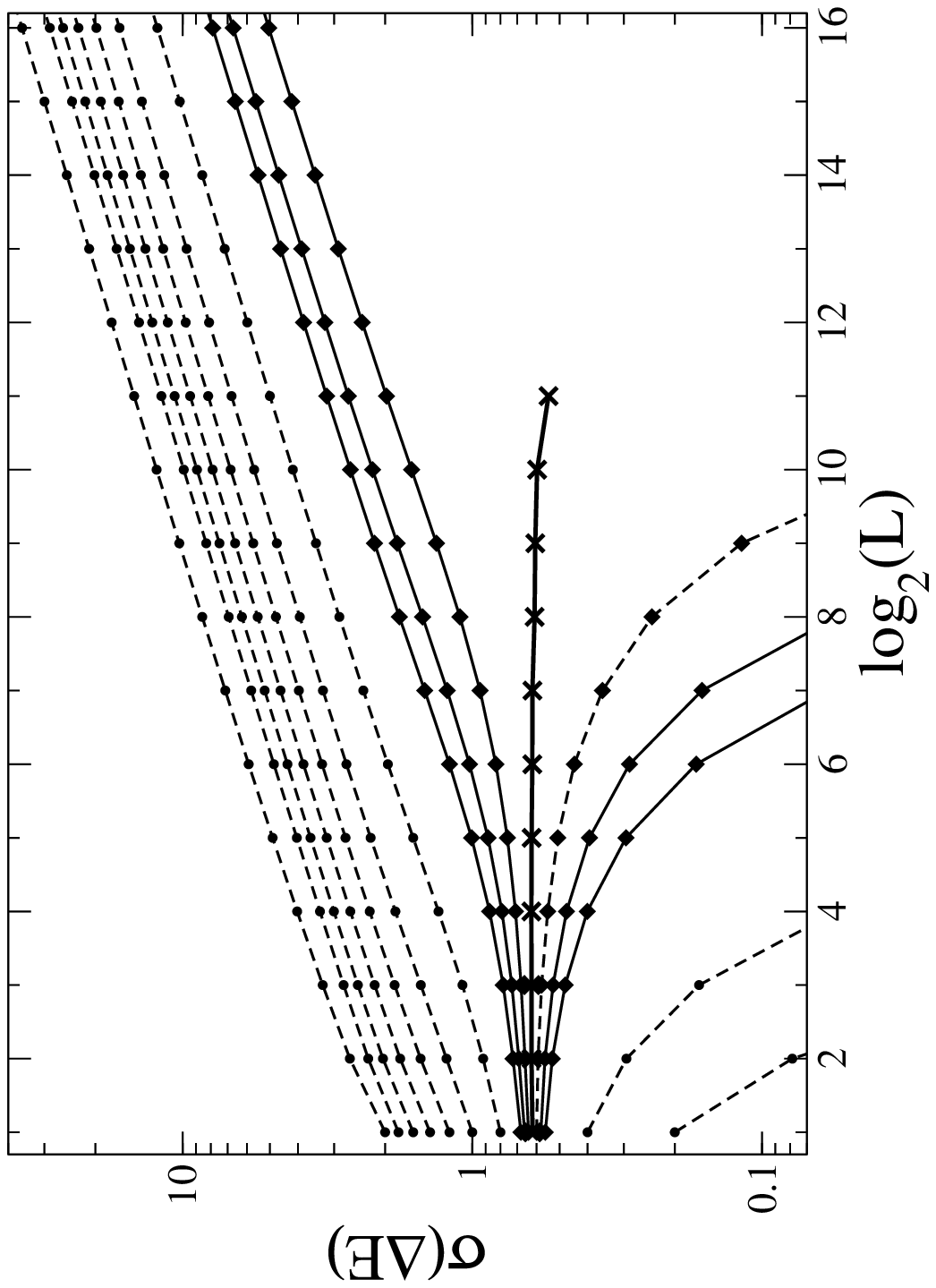} \includegraphics{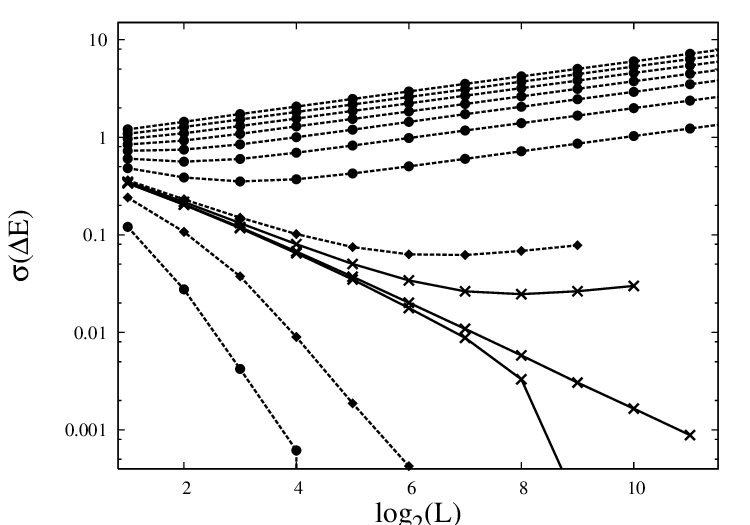}
\caption{Plot of the width $\sigma(\Delta E)$ of the defect energy
distribution as a function of systems size $L$ for various bond
fractions $p$ using discrete bonds (top) and Gaussian bonds (bottom)
in the Migdal-Kadanoff lattice. In each plot, dashed lines from bottom
to top refer to $p=0.1,0.2,\ldots,1$. Solid lines refers to data with
$p=0.28,0.29,\ldots,0.34$ on top, and $p=0.28$, $0.2818$, and $0.29$
on the bottom. In both plots, the data points with circles were
obtained using the $k=1$ implementation from
Ref.~\protect\cite{MKpaper}, those with diamonds using $k=2$, and
those with crosses near the respective $p^*$ required the exact
algorithm. Note that at $p^*\approx0.310>p_c$ on top, the slope is
near vanishing, while for $p^*=p_c=0.2818$ on the bottom, the slope
indicates a well-pronounced power-law decay. }
\label{defectscal}
\end{figure}

Our numerical studies on the hierarchical lattice have been conducted
with the algorithm described in Ref.~\cite{MKpaper}. It is based on
the evolution of bond-pools (see Fig.~\ref{hierlat}) of size $A_I$
from generation $I$ to generation $I+1$, similar to the procedure
already used in Refs.~\cite{BM,ABM}. In our algorithm, though, an
existing bond at generation $I$ is replaced with a new bond (to keep
$A_I$ constant) for every $k$ new bonds that are added at generation
$I-1$. This procedure is legitimate in principle, as even neighboring
sub-elements in the graph act independently and only effect each other
in a collective sense just as represented by the bonds replacing them
in the next generation. The danger is that the diversity in the pool
of those bonds is insufficient, leading to creeping spurious
correlations and difficult-to-perceive drifts away from the true
values of observables. Since the size of the pool of bonds has only a
minor effect the on computational effort for $k=1$, we can re-run the same
calculation repeatedly to a high number of generations using
ever-larger pool sizes $A_I$ until the data becomes insensitive to
$A_I$. We have used this procedure to generate most of the plots in
Figs.~\ref{defectscal}.

Further problems arise when we use this algorithm near the critical
point $p^*$. There, the bond distribution is torn between the trivial
fix point of vanishing width in the paramagnetic phase and the true
glassy state with subtle, long-rang correlations between bonds of the
previous and following generations, and small fluctuations can severely
bias the evolution. At that point, we have to resort to slower
regeneration rates with $k>1$, or even the exact algorithm for which
$k=2^d$.

Our numerical studies, as shown in Fig.~\ref{defectscal}, confirm the
picture described in Sec.~\ref{defect}. To induce an interface in the
MK as described in the Introduction we consider only the left-most and
the right-most spin in each graph as entire boundary. This is
meaningful, since each of these spins connects to $O(L)$ other spins,
see Sec.~\ref{MKlat}. Then, the defect energy of a lattice at
generation $I-1$ (size $L=2^{I-1}$) is simply (twice) the value of the
bond of generation $I$ replacing that graph (see
Fig.~\ref{hierlat}). In this sense, the defect energy $\sigma(\Delta
E)$ can be interpreted as an effective coupling between both sides of
the defect interface~\cite{BBF}; if that coupling strengthens with
distance $L$, the system is in an ordered state, and vice versa.

Note that in Fig.~\ref{defectscal} for $p<p^*$ the data evolves
towards the $p=0$ fix-point with $\sigma(\Delta E)=0$, while for
$p>p^*$ it invariably evolves to the $p=1$ fix-point with scaling
behavior $\sigma(\Delta E)\sim L^y$. Fig.~\ref{defectscal} clearly
suggests that $p^*=p_c=0.2818$ for the continuous Gaussian bond
distribution, while the discrete $\pm J$ bond distribution favors
$p^*=0.3103>p_c$. At $p^*$, $\sigma$ varies distinctively different with
$L$ than even for bond densities $p$ quite near to $p^*$. In
particular, the behavior of $\sigma$ at the respective $p^*$ appears
to confirm the predictions for the scaling in Eq.~(\ref{yPeq}) with
$y_P=0$ (i.~e. $\phi=0$) for discrete bonds, and a non-trivial
exponent we measure to be about $y_P=-0.9(1)$ for Gaussian bonds.

In the following, we want to test the predictive power of the
finite-size scaling Ansatz in Eq.~(\ref{fiteq}) as argued on the basis
of Eq.~(\ref{extendedyeq}). This Ansatz has been applied to the numerical
measurements of the glassy state ($p>p^*$) on large, bond-diluted lattices
as recently proposed in Refs.~\cite{stiff1,stiff2}. In this scheme, the
properties of the zero-temperature fix-point are determined for a
number of intermediate bond-densities $p$ -- sufficiently above $p^*$
to be glassy and sufficiently small to achieve large system sizes $L$.
In particular, Refs.~\cite{stiff1,stiff2} focused on the defect energy for
finite dimensional lattices. This procedure should be applicable also to
other observables.

\begin{figure}
\vskip 2.2in \includegraphics{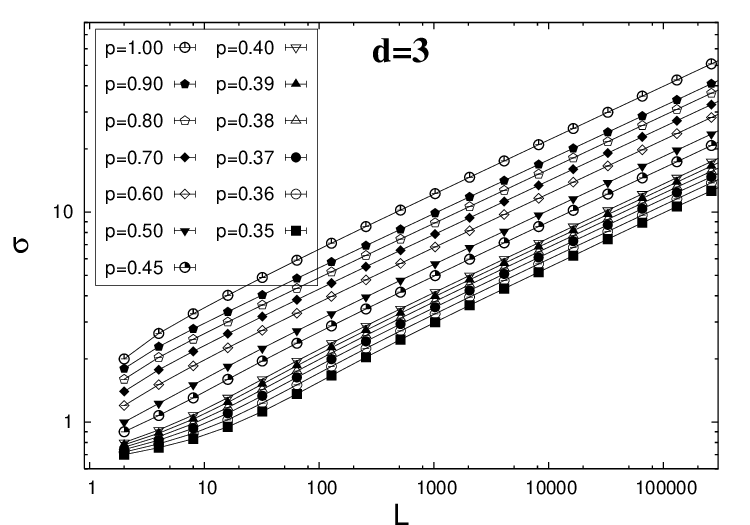}
\caption{Plot of the raw data for $\sigma(\Delta E)$ as a function of
systems size $L$ for bond fractions $p^*<p\leq1$ using discrete $\pm
J$-bonds. Some of this data is already shown in
Fig.~\protect\ref{defectscal}. Of note is that there appears to be no
variation with $p$ in the asymptotic scaling, that scaling corrections
at small $L$ are least noticeable for intermediate $p$, and that there
is an anomalous gap between data for $p=0.9$ and $p=1$.}
\label{pmJraw}
\end{figure}

First, we consider $\pm J$ bond-distributions, which had been used
exclusively in Refs.~\cite{stiff1,stiff2}. In Fig.~\ref{pmJraw} we
plot without any scaling all data we have obtained with the method
above for $p=0.35,\ldots,1$. It's worth mentioning two features of that
data: (1) scaling corrections for smaller $L$ appear to change sign at
around $p\approx0.45$, suggesting that those corrections are weakest
at intermediate values of $p$ instead of at $p=1$, against expectation. 
(2) The data for $p\to1$ initially narrows (on this
logarithmic scale) for equal increments ($\Delta p=0.1$), but then
exhibits an increased gap in the jump from $p=0.9$ to $p=1$. Both of
these features have also been observed for the EA data in
$d=3$~\cite{stiff1,stiff2}. The first feature seems insignificant here
for data extending out to $L>10^5$, but the suppression of scaling
corrections becomes extremely helpful when the maximal attainable $L$
is small, as for the EA model. This fact has also been exploited
in Ref.~\cite{Jorg}. The second feature could be explained with the
requirement for the scaling Ansatz of $\xi\sim(p-p^*)^{-\nu^*}\gg1$,
which may not be satisfied for any $p$ too far from $p^*$. In
Ref.~\cite{MKpaper}, similar rapid variations in an observable (the overlap)
for $p\to1$ have been observed, although a connection to the
variations in the amplitude of $\sigma$ here is not clear.

For this discrete bond-distributions, the scaling Ansatz in
Eq.~(\ref{extendedyeq}) simplifies, since we can assume $y_P=0$, see top of
Fig.~\ref{defectscal}. Hence, we fit the data obtained for various
bond densities $p>p^*$ in Fig.~\ref{pmJraw} directly to the form proposed in
Eq.~(\ref{fiteq}), where the scaling variable $x$ is adjusted to
provide the best data collapse. The unknown scaling function $f(x)$ has
been replace by a fitting constant, $f(\infty)$, to capture its
leading asymptotic behavior for large $L$.

\begin{figure}
\vskip 2.2in \includegraphics{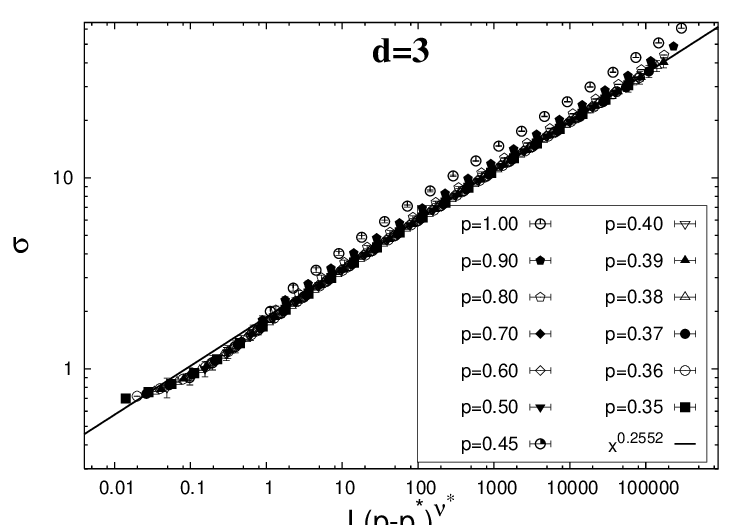}
\caption{Collapse of the data from Fig.~\protect\ref{pmJraw} according
  to the scaling Ansatz in Eq.~(\protect\ref{fiteq}), but using the
  exactly-known values for $p^*$, $\nu^*$, and $y=0.2552$. Here, only
  $f(\infty)$, an overall amplitude, remains as a free fitting parameter.}
\label{pmJfixed}
\end{figure}

\begin{figure}
\vskip 2.2in \includegraphics{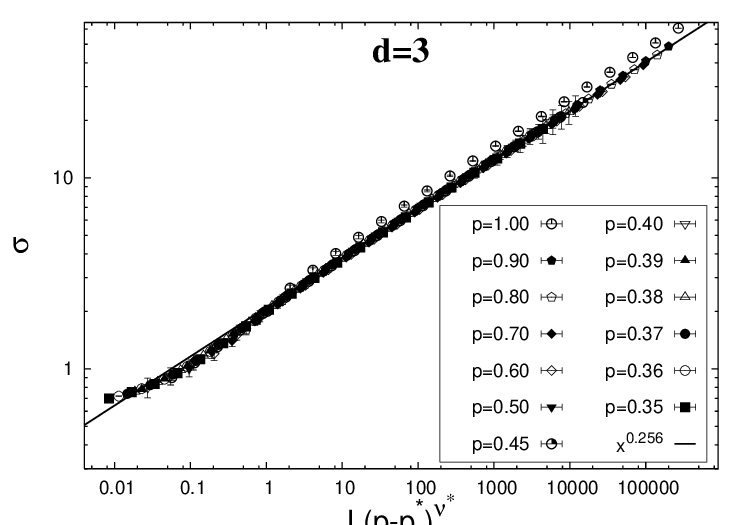}
\caption{Collapse of the data from Fig.~\protect\ref{pmJraw} according
  to the scaling Ansatz in Eq.~(\protect\ref{fiteq}). Here, the values
  for $f(x)$, $p^*$, $\nu^*$, and $y$ are all determined from the fit
  of data in the scaling regime only, although all data is displayed.}
\label{pmJbest}
\end{figure}

The advantage of the hierarchical lattice is that we already know the
expected values for the parameters involved in such a fit,
i.~e. $p^*=0.31032$, $\nu^*=1.5373$~\cite{BF}, and
$y\approx0.2552$~\cite{ABM,MKpaper}. Thus, first we can attempt to
utilize this knowledge to collapse the data in Fig.~\ref{pmJraw} by
fixing $p^*$, $\nu^*$, and $y$, only leaving $f(\infty)$ to be
fitted. As can be seen in Fig.~\ref{pmJfixed}, the data collapses
exceedingly well even for small $x$, as long as $p$ is not too
large. Scaling ensues quickly for $x>1$, which would allow for an
excellent fit, {\it if} one were to exclude data for
$p>0.7$. Remarkable is the quality of the collapse for data reaching
below $x<1$. It reveals a concave shape for $f(x)$, which explains the
finite-size corrections in Fig.~\ref{pmJraw}: Data with lower $p$
first rises slowly, than more rapidly for increasing $L$, before
scaling settles in. In turn, data with higher $p$ immediately rises
rapidly before settling into a slower asymptotic growth. Data with $p$
near unity rises even more rapidly to overshoot the collapse in the
scaling regime.

Clearly, these observations about the intricacies of $f(x)$ are far
too subtle to be of any use in the more typical situations where we
have no prior knowledge of the parameters. Worse yet, even with that
knowledge, corrections due to the finite size of
$\xi=(p-p^*)^{-\nu^*}$ in particular appear to prevent a collapse of
the data in much of the asymptotic scaling regime. Yet, it seems
obvious that the data for {\it all} $p>p^*$ and large enough $L$
exhibits scaling according to Eq.~(\ref{yeq}), providing information
we desire to exploit. The precise value of the parameters $p^*$ and
$\nu^*$ are important for the behavior of $f(x)$ near $x\sim1$, but
have little effect on the scaling behavior for $x\gg1$. Thus, instead
of correcting the fit to extract accurate values for all parameters,
we cut data that does not appear to scale well. Then, we can collapse
the data in the scaling regime according to Eq.~(\ref{fiteq}), merely
using $p^*$, $\nu^*$, and $f(\infty)$ as free parameters to facilitate
an accurate determination of $y$ only. 

{}Fig.~\ref{pmJbest} displays the same data as before, but only data
judged by inspection to be sufficiently scaling ($L>1024$ at $p=0.35$
down to $L>8$ for $p=0.9$, see Fig.~\ref{pmJraw}) has been used for
the fit.  {\it All} data for $p=1$ has been explicitly excluded due to
the anomalous jump in the amplitude of $\sigma$ noted in
Fig.~\ref{pmJraw}. These choices are reflected in the collapse: All
but $p=1$-data combines exceedingly well for $x\gg1$, predicting 
$y=0.256(1)$, only 1/2\% above the exact value. For data
scaling over almost 5 decades, one may have expected more accuracy for
$y$, but slowly decaying scaling corrections~\cite{MKpaper}, and the
smallness of $y$ itself, limit the relative accuracy. The value for
$y$ remains quite robust under changes in the data points included in or
excluded from the fit. On the other
hand, the fit selects $p^*\approx0.29$, $\nu^*\approx1.94$, and
$f(\infty)\approx2.1$. Note that the fitted value for $\nu^*$ is
larger than the actual value. It is this increase in $\nu^*$ that
facilitates the collapse of the data in the asymptotic regime,
$x\gg1$. Consequently, the collapse of the data excluded from the fit
at $x\leq1$ is somewhat poor (but still not too bad here). This is
exactly the approach adopted in Refs.~\cite{stiff1,stiff2} to extract
the best possible estimate for $y$. Hence, as noted there, the values
for any parameter aside from $y$ fitted in this way must be treated
with caution.

\begin{figure}
\vskip 2.2in \includegraphics{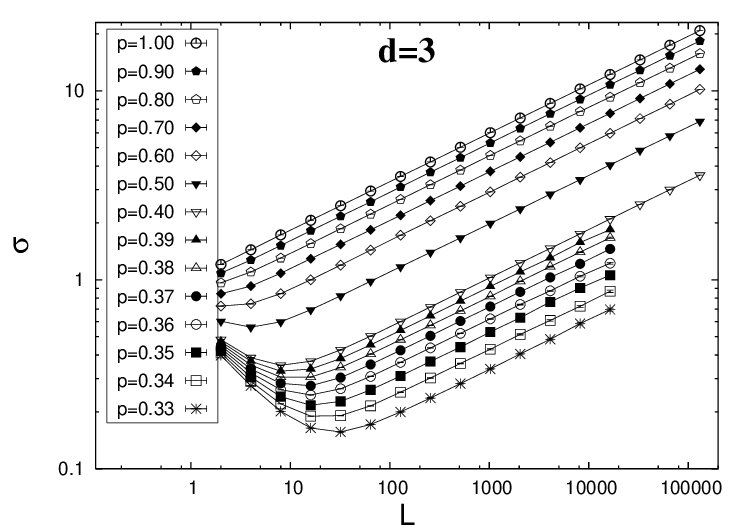}
\caption{Plot of the raw data for $\sigma(\Delta E)$ as a function of
systems size $L$ for bond fractions $p^*<p\leq1$ using continuous
Gaussian-distributed bonds. Some of this data is already shown in
Fig.~\protect\ref{defectscal}.}
\label{Gaussraw}
\end{figure}

\begin{figure}
\vskip 2.2in \includegraphics{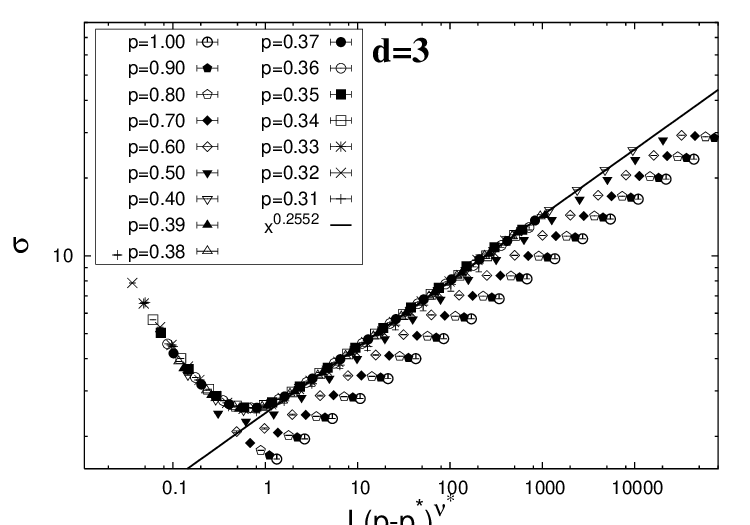}
\caption{Collapse of the data from Fig.~\protect\ref{Gaussraw}
  according to the scaling Ansatz in Eq.~(\protect\ref{fitphieq}), but
  using the exactly-known values for $p^*$ and $\nu^*$ (then $y$ is
  not needed). Here, only $f(\infty)$ and $\phi$ remain as free
  fitting parameters.}
\label{Gaussfixed}
\end{figure}

\begin{figure}
\vskip 2.2in \includegraphics{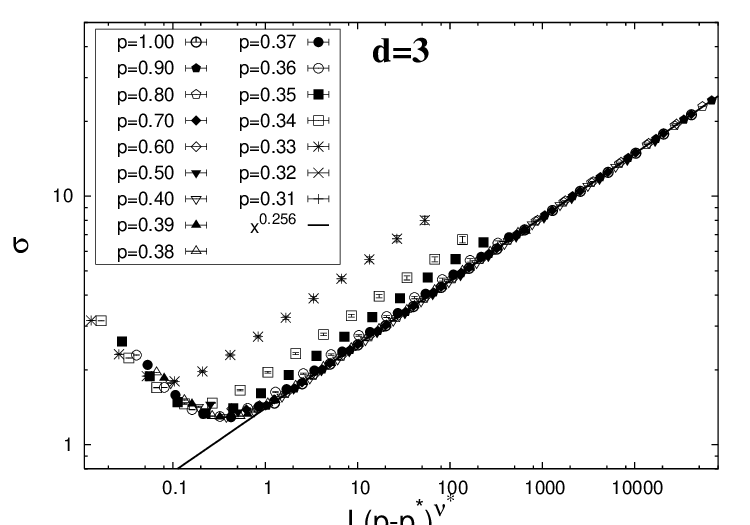}
\caption{Collapse of the data from Fig.~\protect\ref{Gaussraw} according
  to the scaling Ansatz in Eq.~(\protect\ref{fitphieq}). Here, the values
  for $f(x)$, $p^*$, $\nu^*$, $\phi$ and $y$ are all determined from the fit
  of data in the scaling regime only.}
\label{Gaussbest}
\end{figure}

We can proceed in a similar manner for the bond-diluted hierarchical
lattice with a continuous Gaussian bond distribution. First, we plot
the raw data obtained with the algorithm described above in
Fig.~\ref{Gaussraw}. In this case, finite-size corrections are more
pronounced but appear to diminish more gradually towards $p=1$. The
scaling arguments from Sec.~\ref{defect} would suggest to fit the data
to a form derived from Eq.~(\ref{extendedyeq}),
\begin{eqnarray}
\frac{\sigma(\Delta E)_{L,p}}{\left(p-p^*\right)^{\phi}}\sim
f\left(\infty\right)~x^y.
\label{fitphieq}
\end{eqnarray}
Here, it appears that the scaling collapse involves yet another
parameter, the thermal-percolative cross-over exponent $\phi$ for the
scaling variable $x=L/\xi(p)$ at $p\to p^*=p_c$.  In contrast to $p^*$
and $\nu^*$, which can be determined analytically for the hierarchical
lattice, $\phi$ is similarly nontrivial as $y$ itself, and is even
harder to estimate numerically. By definition, $\phi=-\nu y_P$ and
$y_P$ has to be obtained exactly at $p=p^*$, see Eq.~(\ref{yPeq}),
where the algorithm is the most delicate, limiting us to
$L\leq2^{11}$. As shown in Fig.~\ref{defectscal} (bottom), from the
data for $\sigma$ at $p^*$ we can extract $y_P=-0.9(1)$. With the
percolation exponent, $\nu=1.22$, we obtain $\phi=1.1(1)$.

In Fig.~\ref{Gaussfixed}, we collapse the data from
Fig.~\ref{Gaussraw} using Eq.~(\ref{fitphieq}) by fitting $f(\infty)$
and $\phi$, but holding $p^*$ and $\nu^*$, and $y$ fixed. Similar to
Fig.~\ref{pmJfixed}, the collapse proceeds well for data that is near
$x\sim1$ and has $p$ near $p^*$. Already for intermediate values of
$p$, the data spreads widely, suggesting that $\xi(p)$ is too small
there. We obtain a fitted value of $\phi\approx0.9$, not too far from
the determination via $y_P$ above.

Again, the focus on properties associated with $p^*$ gave us a good
collapse near $x\sim1$, but not about the desired asymptotic scaling
regime for $y$ at $x\gg1$. Consequently, we again exclude all data in
the fit that is by inspection of Fig.~\ref{Gaussraw} not yet in that
asymptotic regime and proceed with an unrestricted fit involving all
parameters. The result of such a collapse is shown in
Fig.~\ref{Gaussbest}. Now the data collapse is excellent in the
asymptotic regime for data of sufficiently large $L$ and large enough
$p$, including $p=1$, but poor near $p=p^*$ and for $x\lesssim1$ for
the excluded data. The fitted values here are $p^*\approx0.32$, $\nu^*\approx1.20$,
$\phi\approx0.51$, and $y\approx0.256$, again within 1/2\%
error of the value directly determined at $p=1$. Surprisingly, even $\nu^*$ is well-approximated now while only
$\phi$ is substantially off. In fact, ignoring the correction in the
scaling behavior due to the exponent $\phi$ (i.~e. $\phi=0$), the data collapse
proceeds even more favorably, giving $y=0.2557$ but $p^*=0.31$ and
$\nu^*\approx3$. Having one less free parameter makes the data
collapse a bit more robust, with $\nu^*$ picking up the error due to
a less adequate scaling Ansatz. That a collapse of the data
succeeds even when we ignore $\phi$ is a
reflection of the fact that in the limit of large argument for $f$,
only two of the three exponents $y$, $\phi$, and $\nu^*$ are
independent. Eq.~(\ref{fitphieq}) reduces to $\sigma\sim
f(\infty)L^y(p-p^*)^{\phi+y\nu^*}$, and setting  $\phi=0$ leads us
right back to Eq.~(\ref{fiteq}).

\section{Edwards-Anderson Model with Gaussian Bonds}
\label{Gauss}
We have already pointed out the striking similarities between the data
for defect energies on $d=3$ hierarchical lattices here and on cubic
lattices in Ref.~\cite{stiff1,stiff2} for $\pm J$ bonds. We complement
this comparison here with a study of the EA on cubic lattices with
Gaussian bonds. Such a study allows us to probe some of the assertions
leading to Eq.~(\ref{fiteq}). It also tests the universality of $y$
with respect to the details of the bond distribution. The well-known
result for $d=1$ mentioned in the Introduction and recent studies for
$d>1$~\cite{HY,AMMP,BKM} have show that there are significant
differences in the results for the stiffness exponent with respect to
bond distribution below the lower critical dimension, where $y<0$,
while Ref.~\cite{BKM} has argued that universality should hold when
$y>0$. Our findings here, and for the hierarchical lattice
above, support this point.

\begin{figure}
\vskip 2.2in \includegraphics{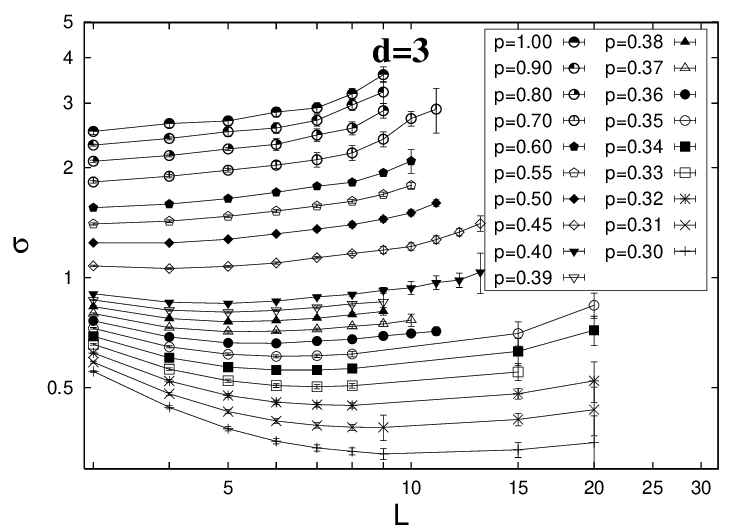}
\caption{Plot of the raw data for $\sigma(\Delta E)$ as a function of
systems size $L$ for bond fractions $p_c=p^*<p\leq1$ using continuous
Gaussian-distributed bonds on a cubic lattice. At nearly each $p$,
long transients are followed by an intermediate scaling regime, cut
short at larger $L$, when numerical inaccuracies result in a
systematic drift.}
\label{Gauss3draw}
\end{figure}

\begin{figure}
\vskip 2.2in \includegraphics{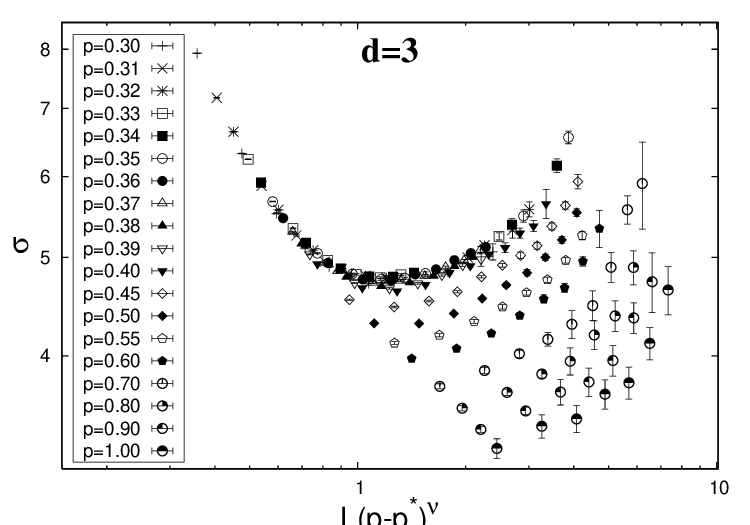}
\caption{Collapse of the data from Fig.~\protect\ref{Gauss3draw}
  according to the scaling Ansatz in Eq.~(\protect\ref{fitphieq}). The
  values for the parameters had to be fixed by hand at
  $p^*=p_c\approx0.248$, $\nu^*\approx0.75$, and $\phi=0.9$.}
\label{Gauss3dfixed}
\end{figure}

\begin{figure}
\vskip 2.2in \includegraphics{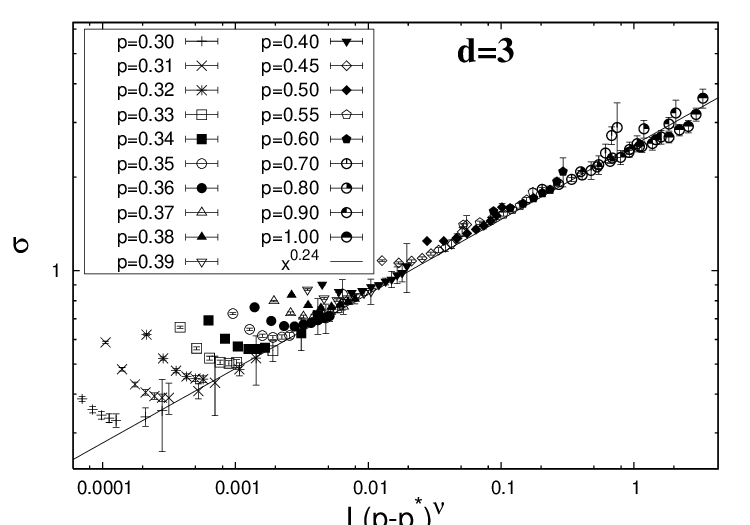}
\caption{Collapse of the data from Fig.~\protect\ref{Gaussraw}
  according to the scaling Ansatz in Eq.~(\protect\ref{fiteq}),
  i.~e. $\phi=0$. Here, the values for $f(\infty)$, $p^*$, $\nu^*$,
  $y$ are all determined from the fit of data in the scaling regime
  only, although all data from Fig.~\protect\ref{Gaussraw} are
  shown. Despite the obvious shortcomings of the data, the fitted
  value of $y\approx0.23$ compares reasonably well with the best-known
  value of $y=0.24(1)$~\protect\cite{stiff1,stiff2}, drawn here for
  reference (line).}
\label{Gauss3dbest}
\end{figure}

Unlike for hierarchical lattices, the numerical effort require to
achieve any reasonable system size $L$ in the determination of defect
energies on cubic lattices grows exponentially with the number of
variables; the problem is NP-hard~\cite{Barahona}. While the method of
reducing low-connected variables is applicable just as well for the
Gaussian distribution, our implementation of the extremal optimization
heuristic does {\it not} perform well for these systems. Although it
might be capable to yield reasonably accurate predictions for ground
states themselves, we notice a significant drift in the data for the
defect energy, formed out of the difference between two closely
related ground states, starting with remainder graphs of size
$n\geq250$. As Fig.~\ref{Gauss3draw} shows, much of the obtained data,
both for small and larger $L$, is not scaling and has to be
discarded. We note that the long transients even before any asymptotic
scaling is reached for small $L$ is again reminiscent of the
hierarchical lattice as in Fig.~\ref{Gaussraw}.

An attempt to collapse the data according to Eq.~(\ref{fitphieq})
as before requires the additional effort in determining $y_P$ from
Eq.~(\ref{yPeq}). A determination of the defect energy at $p_c$ has
the advantage that the exact reduction method almost always succeeds
completely, obviating the application of any optimization and large
lattice sizes can be reached~\cite{Marchetti}. Our preliminary studies
for systems up to $L=100$ have yielded an exponent of $y_P=-1.28(2)$.
The consensus of results for the correlation exponent in $d=3$ for
percolation seems to be $\nu=0.86(2)$~\cite{Martins}, which results in
a thermal-to-percolative crossover exponent, see Eq.~(\ref{Tgeq}), of
$\phi=1.10(4)$. Note that while $\nu<1$ and $|y_P|>1$ here, opposite
to the hierarchical lattice in Sec.~\ref{numerics}, the values of
$\phi$ are indistinguishable within errors.

Similar to Fig.~\ref{Gaussfixed}, one may try to collapse the data
according to the scaling Ansatz in Eq.~(\ref{fitphieq}) by fixing
$p^*$, $\nu^*$ and $y$ to their best-known values and fitting for
$f(\infty)$ and $\phi$. Such a fit does not converge. We managed only
to collapse the data for $p\to p^*$ by hand with the best-known value
for $p^*=p_c\approx0.248$~\cite{Ziff}, but $\nu^*\approx0.75$, and
$\phi=0.9$, somewhat below their best-known values. The result in
Fig.~\ref{Gauss3dfixed}, while far less convincing for this limited
data set, parallels that in Fig.~\ref{Gaussfixed} to a large extend.

Finally, we set  $\phi=0$, which reverts Eq.~(\ref{fitphieq})
into Eq.~(\ref{fiteq}), and eliminate all data that is obviously not
scaling. Pursuing a fit
of the remaining data according to Eq.~(\ref{fiteq}) in the asymptotic
regime, $x\gg1$, without fixing any parameters, Fig.~\ref{Gauss3dbest}
is obtained. The result is noisy and apparently  unsatisfactory: The
value of $\nu\approx3.1$ is as large for Fig.~\ref{Gaussbest}, which
separates this data in the scaling regime for each $p$ into almost
disconnected groups. Yet, the fit obtains a value of
$y\approx0.23$, in reasonable agreement with the best-known value of
$y=0.24(1)$~\cite{stiff1,stiff2}.

Clearly, as for the hierarchical lattice at the end of
Sec.~\ref{numerics}, the focus on data with $x\gg1$, i.~e. $p$
sufficiently larger than $p^*$ such that $L\gg\xi(p)$, makes the
details of the transition at $p\to p^*$ (such as the value of $\phi$)
irrelevant for the determination of $y$. This suggests
Eq.~(\ref{fiteq}) as the appropriate scaling Ansatz to extract $y$, as
proposed in Refs.~\cite{stiff1,stiff2}, even for Gaussian bonds.

\section{Conclusions}
\label{conclusion}
We have explored a recently proposed method of collapsing data
obtained on bond-diluted lattices to estimate low-temperature scaling
properties for the Edwards-Anderson model. Using bond-diluted
hierarchical lattices from the Migdal-Kadanoff bond-moving scheme in
$d=3$, for which many properties at $T=0$ are known exactly or with
high accuracy, the validity of the method is probed. The data obtained
for the defect energy of the hierarchical lattice proves to exhibit
the same qualitative features as the -- much more limited -- data for
the EA, respectively for discrete and continuous bond disorder.

The scaling Ansatz used to collapse the data, most generally
Eq.~(\ref{extendedyeq}), was proposed in the neighborhood of a $T=0$
phase transition at bond density $p^*$ between paramagnetic and spin
glass behavior, which is closely associated with the bond percolation
transition at $p_c$~\cite{BBF,BF}. Eq.~(\ref{extendedyeq}) requires
both, the divergence of the system size $L$ {\it and} of the
correlation length $\xi=(p-p^*)^{-\nu^*}$.  In contrast, the desired
scaling behavior is associated with a $T=0$ fixed-point characteristic
of the entire spin glass regime for $L\gg1$, independent of $\xi(p)$
as long as $L\gg\xi(p)$. In particular, valuable data obtained for
large $L$ but intermediate values of $p$, for which $\xi(p)$ remains
small, would have to be discarded in an Ansatz based on $p\to
p^*$. Thus, a naive Ansatz, Eq.~(\ref{fiteq}), based on the limit
$L\gg\xi(p)$ (i.~e. $x\to\infty$) in Eq.~(\ref{extendedyeq}), exploits
the obtained numerical data optimally (after data with $x\lesssim1$ is
cut). This Ansatz is ``naive'' in the sense that $x=L(p-p^*)^{\nu^*}$
is not a true scaling variable, and the obtained values of the fitted
parameters do not correspond to those defined by
Eq.~(\ref{extendedyeq}). Those parameters provide useful degrees of
freedom in the fit to remedy unknown corrections in $\xi(p)$, as our
discussion here shows. In fact, Eq.~(\ref{fiteq}) has been used
successfully in Refs.~\cite{stiff1,stiff2,BH}, where very robust
scaling behavior was extracted for the desired defect scaling exponent
$y$, but unrecognizable values have been found for $p^*$ and
$\nu^*$. Unknown corrections in particular prevented a data collapse
for data at $p=1$ there, a fact closely mirrored by the hierarchical
lattice with a discrete bond distributions here, see
Fig.~\ref{pmJraw}.

Clearly, the system sizes $L$ attainable for the hierarchical lattice
are unrealistic for the EA, and the unknown corrections to scaling
disfavor any finite size scaling Ansatz in comparison to the accuracy
obtained in a simple extrapolation of $p=1$
data~\cite{ABM,MKpaper}. Yet, the striking similarities in the
behavior of the data for defect energies of the hierarchical lattice
and EA should be noted. The Gaussian bond distribution results in
extended transient behavior until scaling is reached, making it
impractical to extract the stiffness exponent $y$ for the EA, even if
an optimization heuristic would be applied that could handle remainder
graphs beyond the limit of $\approx250$ spins used here. Conversely,
the cross-over at intermediate $p$ for system size corrections
minimizes transient behavior for $\pm J$ bonds, leading to the best
data collapse for values of $L$ realistic achievable for the EA. This
intermediate window in $p$ is invariant but narrows for $d\to\infty$,
allowing for reasonable determinations of $y_d$ for as high a
dimension as $d=7$, with scaling in the data collapse extending over
two decades in $d=3$ to merely half a decade in
$d=7$~\cite{stiff2}. These new values for $y_d$ allow for a direct
comparison with mean field predictions~\cite{SK_EO}. Our study here
should add some confidence into those findings.

\section*{Acknowledgments}
SB would like to that A. Bray and M. Moore for helpful
discussions. This work was supported by NSF grant DMR-0312510.

\end{document}